\documentclass[aps,prper,twocolumn,superscriptaddress,letterpaper,longbibliography]{revtex4-1}

\usepackage{graphicx}
\usepackage{mdframed}
\usepackage{framed}
\usepackage{indentfirst}
\usepackage{natbib}
\usepackage{amsmath,amssymb}
\usepackage{subfigure}
\usepackage{enumitem}
\usepackage{multirow}
\usepackage{epstopdf}

\begin{document}

\title{Investigating students' behavior and performance in online conceptual assessment}



\author{Bethany R. Wilcox}
\affiliation{Department of Physics, University of Colorado, 390 UCB, Boulder, CO 80309}

\author{Steven J. Pollock}
\affiliation{Department of Physics, University of Colorado, 390 UCB, Boulder, CO 80309}

\begin{abstract}
Historically, the implementation of research-based assessments (RBAs) has been a driver of educational change within physics and helped motivate adoption of interactive engagement pedagogies.  Until recently, RBAs were given to students exclusively on paper and in-class; however, this approach has important drawbacks including decentralized data collection and the need to sacrifice class time.  Recently, some RBAs have been moved to online platforms to address these limitations.  Yet, online RBAs present new concerns such as student participation rates, test security, and students' use of outside resources.  Here, we report on a study addressing these concerns in both upper-division and lower-division undergraduate physics courses.  We gave RBAs to courses at five institutions; the RBAs were hosted online and featured embedded JavaScript code which collected information on students' behaviors (e.g., copying text, printing).  With these data, we examine the prevalence of these behaviors, and their correlation with students' scores, to determine if online and paper-based RBAs are comparable.  We find that browser loss of focus is the most common online behavior while copying and printing events were rarer.  We found that correlations between these behaviors and student performance varied significantly between introductory and upper-division student populations, particularly with respect to the impact of students copying text in order to utilize internet resources.  However, while the majority of students engaged in one or more of the targeted online behaviors, we found that, for our sample, none of these behaviors resulted in a significant change in the population's average performance that would threaten our ability to interpret this performance or compare it to paper-based implementations of the RBA.
\end{abstract}

\maketitle

\section{\label{sec:intro}Introduction and Motivation}

Research-based assessments (RBAs) have become a cornerstone of physics education research (PER) due in large part to their ability to provide a standardized measure of students' learning that can be compared across different learning environments or curricula \cite{madsen2017rbaRL}.  As such, these assessments are a critical step along the path towards making evidenced-based decisions with respect to teaching and student learning.  RBAs have historically been a strong driver in promoting the need for, and adoption of, educational reforms in undergraduate physics courses (e.g., \cite{beichner2007scaleup, hestenes1987modeling, crouch2001pi}).  It can be argued that, without the invention and consistent use of RBAs, the PER community might not have the same focus on active learning and interactive engagement that is does today.  

However, despite their value, there are a number of barriers to wide-scale implementation of RBAs that stand in the way of their integration into physics departments \cite{wilcox2016admin, macisaac2002standardized}.  For example, most of the existing RBAs require that an instructor sacrifices 1-2 full class periods to administering the RBA pre- and post-instruction.  For many instructors feeling pressure to cover as much content as possible over the course of a semester, this sacrifice is difficult to justify.  In addition to the demand for class time, instructors must also sacrifice valuable time outside of class to analyze their students' performance.  Many instructors are not experts in assessment and struggle with analysis and interpretation of their students' scores.  This can make faculty particularly reluctant to sacrifice class time to an assessment that they are ultimately unable to identify actionable results from.  

Recently, physics education researchers have attempted to address both of these challenges by shifting RBAs to online platforms (e.g., \cite{wilcox2016admin, vandusen2015lasso, PhysPortwebsite, holmes2019plic}).  Hosting the RBAs online allows instructors to assign the RBA for student to complete outside of class, freeing them from the need to sacrifice class time.  Moreover, the online platform allows for easy standardization and centralization of the data collection and analysis process.  This has two major advantage for the instructor.  By automating the analysis of students' responses, these centralized systems make is so that the instructor no longer needs to perform this analysis themselves.  Moreover, centralizing data collection ensures the aggregation of comparison data that can be used to facilitate meaningful comparisons and can help instructors, and researchers, to identify actionable implications from their students' performance.  However, while these online systems have a lot of potential for encouraging more wide-spread use of RBAs by removing barriers to their use, these systems bring with them an number of other concerns, particularly around the potential for reduced participation rates, students' use of outside resources, potential for distraction and breaches of test security \cite{macisaac2002standardized}.  

Here, we build on prior work to investigate the extent to which these concerns factor into students' scores when completing standardized physics conceptual assessments online.  We include data from both introductory and upper-level contexts as there are significant differences between the student populations at these two levels, which could have implications for how students engage with an online assessment.  In the next section (Sec.~\ref{sec:background}), we describe prior work around online conceptual assessments.  We then discuss the context and methods used in this study (Sec.\ \ref{sec:context}), and present our findings with respect to students' online behaviors when taking the RBAs as well as how these behaviors correlate with their overall performance (Sec.\ \ref{sec:results}).  Finally, we end with a discussion of our conclusions, limitations, and implications of the study (Sec.\ \ref{sec:discussion}).  

\section{\label{sec:background}Background}

Significant prior work has been done to address some of the concerns around online conceptual assessment as part of the Learning Assistant Student Supported Outcomes (LASSO) study.  Specifically, they have investigated concerns about changes in scores and participation rates between online and paper-based administrations of the Force Concept Inventory (FCI) \cite{hestenes1992fci} and Conceptual Survey of Electricity and Magnetism (CSEM) \cite{maloney2001csem} in the context of introductory courses.  They found that, when looking at all courses in aggregate, participation rates tended to be lower for online RBAs \cite{jariwala2017lasso}.  However, this difference between the two formats vanished when best practices were used for the online implementations.  The best practices identified in the LASSO study include: multiple email and in-class reminders, and offering participation credit for completing both the pre- and post-tests.  Moreover, they also found that, when participation rates were similar, students' overall performance was also statistically comparable \cite{nissen2017lasso}.  

Historically, multiple researchers have address the issue of the comparability of online assessments both within and outside of PER.  For example, MacIsaac \emph{et. al} \cite{macisaac2002standardized} also found no difference in students scores on the FCI between web-based and paper-based administrations.  In addition to investigating overall score, they also saw no difference in performance on individual items and no difference based on students' gender.  However, while studies withing PER have consistently indicated that there is no difference in performance between online and paper-based RBAs, the results from outside PER are more varied.  Many studies have documented no statistically significant difference between students performance on online multiple-choice tests (e.g., \cite{zandvliet1997cat, ladyshewsky2015online}), while others have reported cases where online tests scored statistically higher or lower than associated paper-based tests (see Refs.~\cite{bugbee1996cbtesting, ladyshewsky2015online} for reviews).  The variation in these studies has led to the recommendation that, while online and paper-based tests \emph{can} be equivalent, it should not be \emph{assumed} that they are equivalent until it has been clearly demonstrated that they actually are \cite{bugbee1996cbtesting}.  

A smaller body of work has focused specifically on investigating the validity of concerns about students' use of outside resources (e.g., the internet or other students) or breaches in test security.  For example, Haney and Clark \cite{haney2007cheating} collected timing and path data from students who took a series of online quizzes over the course of a one semester course.  By analyzing patterns in students responses (e.g., similarities in two students response patterns combined with close timing of the two submissions) to identify likely cases of students collaborating on the assignments.  They found that this type of collaboration increased as the semester went on and students adapted to the online quiz format.  They also asked students to self-report whether they collaborated with others for the quizzes and found students reported collaboration with similar frequencies to what was detected in the response patterns.  

Another study conducted in the context of an introductory astronomy course looked at different online student behaviors when taking an online conceptual assessment \cite{bonham2008online}.  In this study, Bonham \cite{bonham2008online} used JavaScripts and other applets to detect when students engaged in behaviors like printing browser pages, coping or highlighted text, and switching in to other browser windows while taking an online astronomy concept assessment.  They found no instances of students printing pages, and only 6 cases (out of 559) they deemed were probable incidence of students copying text.  Students switching browser windows was more common; however, Bonham argued these events appeared random and were not systematically associated with particular questions.  There were several important limitations to Bonham's study.  In browsers other than Microsoft Explorer, copy events and save events were detected through the proxies of highlighting text and page reloads respectively.  As Bonham noted, highlighting text as a proxy for copying results in many false positives, and there was no discussion of how these behaviors related to performance on the RBA.  Here, we replicate and extend the study by Bonham in the context of physics courses at both the upper-division and introductory levels.

\section{\label{sec:context} Context \& Methods}

Four different physics RBAs were used in this study - two upper-division and two introductory.  The two upper-division RBAs used in this study were the Quantum Mechanics Conceptual Assessment (QMCA) \cite{sadaghiani2015qmca} and the Colorado Upper-division Electrostatics Diagnostic (CUE) \cite{wilcox2015cue}.  Both the QMCA and CUE are multiple-choice or multiple-response assessments targeting content from the first semester in a two-semester sequence in junior-level Quantum Mechanics, and Electricity and Magnetism respectively.  The two introductory assessments used were the Force and Motion Conceptual Evaluation (FMCE) \cite{thornton1998fmce} and the Brief Electricity and Magnetism Assessment (BEMA) \cite{ding2006bema}.  The FMCE and BEMA are both multiple-choice assessments targeting content from the first and second semester, respectively, of a two-semester, calculus-based introductory physics course.  All four RBAs were administered online, using the survey platform Qualtrics, during the final week of the regular semester.  The online versions of the RBAs were designed to mirror the paper versions as faithfully as possible.  For example, each separate page on the paper versions was offered as a separate browser page in the online version.  Students could also navigate freely both forward and backward within the assessment, as they would be able to with a paper exam packet.  

Student responses were collected from 2 introductory courses and 10 upper-division courses at eight institutions.  All eight institutions are four-year universities spanning a range of types including three doctoral-granting institutions classified as very high research, two masters-granting institution (one classified as Hispanic-serving), and three bachelors-granting institutions.  The authors taught two of the upper-division courses, and the remaining instructors volunteered.  In all cases, the instructors offered regular course credit to their students for simply completing the RBA (independent of performance).  In most cases, students received multiple in-class reminders to complete the assessment.  After elimination of responses that were marked as invalid (e.g., due to too many blanks; less than 3\% of non-dublicate responses were identified as invalid), participation rates by course varied from 70\% to 100\% over all 12 courses.  Since the goal of this study is not to contrast courses at the same level, the remainder of the analysis will aggregate all introductory and upper-division students separately.  

The breakdown of overall participation rates is given in Table~\ref{tab:rates}.  These rates are somewhat higher than what has been observed for post-test participation in either paper-based or online RBAs in previous studies.  For example, the LASSO study found for their introductory population an average post-test response rate of 66\% for paper-based RBAs and 50\% for online RBAs \cite{jariwala2017lasso}.  We also have historical participation rates available for three of the upper-division courses and both of the two introductory courses in the data set.  Average historical participation rates for these courses were between 60-85\% for both the upper-division courses and introductory courses.  These same courses saw participation rates between 79-97\% in the current data set, suggesting that the participation rate for these courses actually increased somewhat when the RBA was given online.  The fact that the instructors in our data set offered a meaningful amount of regular course credit for students who participated in the RBA, independent of their performance, likely contributed significantly to this increase in participation.  We do not have consistent access to data on the racial or gender distributions for the students in our data set and thus do not report this breakdown here.  

\begin{table}
\caption{Overall participation rates for the introductory and upper-division populations.  Participation rates (``Rate") represent the percentage of the total course enrollment for which we collected valid responses ($N_{valid}$) in the final data set.  Individual course participation rates varied from 70\% to 100\%.  }\label{tab:rates}
 \begin{ruledtabular}
    \begin{tabular} {l c c c}
		& $N_{valid}$ & $N_{roster}$	& Rate \\
		\hline
		Introductory	&	1287	& 1543	&	83.4\% 	\\
		Upper-division	&	308	& 336	&	90.2\% 	\\
	\end{tabular}
 \end{ruledtabular}
\end{table}

On the first page of the assessment, students were instructed to complete the RBA in one sitting without the use of outside resources such as notes, textbooks, or Google.  To capture students' online behaviors, we embedded JavaScript code into the online prompts to look for instances of students copying text, printing from their browser, or clicking into another browser window.  For the upper-division population, the code only recorded when a student copied text, but did not record what text was copied.  However, for the introductory population, we also collected data on what question the text was copied from.  In all cases, these behaviors were time stamped to determine when each action occurred and how many times each student exhibited that behavior.  This JavaScript code could only detect activities that happen at the browser level; activities at the computer level (e.g., taking a screenshot or clicking into another program) were not recorded by the code.  While such data would be useful, modern browsers nearly all have security features to prevent cookies and scripts in browsers from collecting information on activities happening outside the current browser window.  

For browser print commands (e.g., ``control-p") and copy text commands (e.g., ``control-c"), the primary data collected were when and how often these commands were issued.  Data on browser focus were somewhat more complex.  The code was designed to listen for a change in browser focus, and then record whether the RBA tab was visible 4 seconds after the browser focus event occurred.  This allows for a variety of patterns in focus data as students click in and out of the browser tab, which they sometimes did rapidly and repeatedly.  However, in general, if a student clicks into a new browser tab and stays in that tab for more than 4 seconds, the code would record a single browser focus event and tag it ``hidden."  A ``hidden" browser focus event most often means that the student left the RBA without returning to it within 4 seconds.  Alternatively, if the student clicked into another browser tab and then clicked back into the RBA within 4 seconds (and remained there for more than 4 seconds), the code would record two browser focus events -- one for the click out and one for the click in -- and would tag both as ``visible."   A single ``visible" browser focus event most often means that the student returned to the RBA for more than 4 seconds after having left it for any amount of time.

In addition to the data on students' online behaviors, we collected students' scores on the assessment in order to compare the prevalence of students behaviors with their performance on the RBA.  Total time spent on the assessment was approximated using the time elapsed between the time the student clicked on the link for the RBA and the time they submitted it.  As we discuss in more detail in the next section, this duration is only approximate because it does not account for time the student might have spent away from the RBA (e.g., in another browser tab, or not on their computer).

\section{\label{sec:results}Results}

Here, we examine data on students behaviors on online RBAs to determine how prevalent specific online behaviors were for this population of students.  We also examine correlations between these behaviors and students' performance on the assessments overall.  

\subsection{\label{sec:print}Print Events}

The primary concern associated with students printing or saving RBAs is that these students might publicly post the assessment and thus breach the security of the assessment by making it available to other students.  Because the online RBAs were designed to mirror the paper-based versions, each had 10-15 individual pages that the students would work through to see all questions.  This means that to present a significant threat to the security of the assessment, a student would need to print each page of the assessment separately, and in so doing, would register multiple print commands.  To determine the prevalence of students printing their browser page, we include responses from the full data set, including responses marked as ``invalid'' from, for example, students who did not ultimately submit the RBA.  In this full data set of 1879 student responses, only five (two from the introductory population and three from the upper-division) had recorded print events.  Of these, 3 students, all from the upper-division population, had multiple print commands consistent with having saved all or the majority of the assessment pages.  The remaining 2, both from the introductory population, had only 1-2 distinct print events meaning they could, at most, have saved only a small number of questions.  It could be that after beginning the process of saving the questions, these students realized the process would require saving each page of the assessment individually and gave up.  

Print commands themselves do not necessarily indicate a student who is intending to breach the security of the assessment.  In fact, one of the instructors (SJP) reported interacting with a student during help hours in which the students pulled up screenshots of the assessment which he had taken to study from after the fact.  The student made no attempt to hide the screenshots and was upfront with his motivation for taking the screenshots as a study tool.  Moreover, even if a student did post the RBA prompts online, without corresponding solutions, which were never released to the students, it is not clear that access to the RBA prompts alone actually represents a significant threat to the assessment's security or validity.  Additionally, as is standard for paper-based assessments, the formal names or acronyms for the assessments (e.g., the CUE) were not provided to students in the online versions.  

To test for any immediate security breaches of the assessments, we Googled the prompts for each question on all four RBAs used in this study several weeks after the assessments had closed.  The results of these searches varied significantly for the introductory and upper-division assessments.  For the two upper-division assessments (the CUE and QMCA), there was no indication that the item prompts or their solutions had been uploaded in a way that ranked high in Google's listing.  However, as Google's algorithm can change based on search patterns, it is likely necessary to do this type of verification periodically to ensure no solutions have surfaced.  In several cases, Googling the item prompts pulled up PER publications on the test itself, and some of these publications included supplemental material which contained the grading rubrics for the assessment in one form or another (open-ended or multiple-choice).  It is worth noting that in all cases, these rubrics were buried at the end of a long publication or thesis and not clearly marked, and it is not clear if a student who was unfamiliar with the specific publications (or the nature of academic publication more generally) would be able to locate the rubrics without considerable persistence.  However, this suggests that the greatest threat to the security of the upper-division RBAs in an online format may actually be our own publications combined with the fact that the premier PER publication venue is open access.  

Attempts to Google the prompts to the two introductory RBAs (the FMCE and BEMA), however, yielded very different results.  Searching prompts for items on these assessments pulls up images of the exact prompts from the assessment, and accompanying solutions are available on paid solution sites like Chegg or Course Hero.  Any student with an existing subscription to these sites would like be able to find solutions to the FMCE or BEMA questions with relative ease.  These solutions predate this study, and thus represent breaches of security that occurred previously.  The larger online presence of both the introductory RBA prompts and solutions has at least two possible contributing factors.  First, introductory (and largely non-physics major) students may be more likely to engage in behaviors that facilitate quick completion of online assignments rather than prioritizing deep learning of the material.  Thus, they may be more likely to look for, and share, course materials online.  Second, both the FMCE and BEMA are considerably older and more extensively used than the CUE and QMCA.  It may be that solutions to any RBA will eventually make their way online given sufficient time and use, and that the CUE and QMCA are not old enough or common enough to have achieved a significant online presences.  We will discuss additional implications of this pattern in Sec.~\ref{sec:copy}.  

\begin{table*} 
\caption{Duration and number of sustained browser hidden events in the introductory and upper-division student population.  For reference, the total number of valid responses in the introductory and upper-division data sets was $N=1287$ and $N=308$ respectively. }\label{tab:focus}
 \begin{ruledtabular}
    \begin{tabular} {l c c}
		& Introductory & Upper-division \\
		\hline
		Total number of students with 1 or more focus event	&	562	&	159	\\
		\hspace{2mm} Number of focus events per student	&	Median - 2	(Max - 43)	& Median - 2	(Max - 59)  \\
		\hspace{2mm} Number of students with only 1 focus event	&	219	& 66  	\\
		\hspace{2mm} Number of students with 10 or more focus events	&	91	& 20  	\\
		\hline
		Total number of focus events	&	2860	& 725  	\\
		\hspace{2mm} Duration of focus events	&	Median - 21 sec	(Max - 66.7 hrs)	& Median - 34 sec	(Max - 29.3 hrs)  	\\
		\hspace{2mm} Number of focus events less than 1 min	&	2264	& 479  	\\
		\hspace{2mm} Number of focus events greater than 5 min	&	149	& 70  	\\
	\end{tabular}
 \end{ruledtabular}
\end{table*}

\subsection{\label{sec:focus}Browser Focus Events}

Online RBAs introduce a potential for students to become disengaged from the assessment in a way that is less likely in paper-based administrations.  Loss of browser focus is one proxy for students disengaging from the RBA.  Focus events were the most common events in the data set with roughly half of the students (46\%, $N=562$ of 1287 in introductory; 52\%, $N=159$ of 308 in upper-division) with at least one browser focus event in which their RBA window became hidden for more than 4 seconds.  For these students, we examined trends in the number and duration of browser focus events by grouping them to isolate sustained changes in browser visibility.  In other words, if a students' survey page becomes hidden, how long before it becomes visible again, independent of whether there are additional browser hidden events in between (indicating that the student clicked back into the survey window, but did not remain there for more than 4 seconds)?  Here, we will report median and max duration, as the presence of even a small number of outliers makes the average less meaningful.  Table~\ref{tab:focus} reports information on the number and duration of browser focus events in the dataset.  While Table~\ref{tab:focus} reports data for the introductory and upper-division students separately, the trends are comparable between the two levels.  These trends suggest that a large fraction of students in the data set did click out of the assessment tab one or more times while taking the RBA; however, roughly two-thirds of the time they were away from the RBA for no more than 1 minute and less than 10\% left the assessment for longer than 5 minutes.  Moreover, just over a third of students left the assessment only once.  In our experience implementing assessments like these in in-class environments, this frequency and time-frame is generally comparable to how long a student might ``space out" while taking the RBA during class.  

We also examined whether the appearance or duration of loss of focus events correlated with students' scores on the assessment.  In as much as browser loss of focus could be a proxy for distraction, it might be guessed that students with loss of focus events would score lower than others on the RBA.  Alternatively, if the loss of focus is associated with use of internet resources (see Sec.~\ref{sec:copy}), we would anticipate students with loss of focus events to potentially score higher.  To account for difference in average score between courses in the data set, z-scores calculated relative to the average score for each individual class were used in calculating correlations.  Students with loss of focus events scored higher on average by roughly a quarter of a standard deviation than other student for the introductory RBAs (i.e., a z-score difference of $0.26$) and lower on average by roughly a fifth of a standard deviation for the upper-division RBAs (i.e., a z-score difference of $-0.19$). The difference in performance was statistically significant in the case of the introductory courses (Mann-Whitney U $p=0.001$) though small (Cohen's $d=0.26$), and was not statistically significant for the upper-division population.  Additionally, we examined the Spearman correlation coefficient between the total time students spent with their browser hidden relative to their score on the assessment.  We selected the Spearman correlation because it is less sensitive to the presence of outliers than the other coefficients.  Consistent with the differences in average score, we found a statistically significant, though small, correlation between score and total time away from the assessment tab for introductory students ($r=0.16$, $p=0.0001$) and no significant correlation for the upper-division students ($r=-0.1$, $p=0.2$).   

\subsection{\label{sec:copy}Copy Events}

The primary concern associated with students copying text from an online RBA is that students may do so in order to search the internet in an attempt to ``look up" the correct answer.  Table~\ref{tab:copy} shows the prevalence of copy events within our data set, showing that roughly a tenth of the student in the data set had one or more copy events.  A copy event, on its own, does not necessarily mean that the student was attempting to web search answers to the questions.  However, if a student copies text with the intention of searching the web for that text, this behavior would most likely be characterized by a copy event followed immediately by a sustained browser hidden focus event.  To investigate this, we looked for copy events followed within 5 seconds by a sustained browser loss of focus event and counted how many times this occurred for each student.  We found that more than three quarters of the copy events ($N=654$ of 861 events for introductory, and $N=56$ of 67 events for upper-division) fell into this category.  This indicates that a majority of copy events were immediately followed by the student switching into a new browser window and remaining there for more than 4 seconds, consistent with the pattern we would expect if they were trying to web search the item prompts.  The remaining copy events that were not followed by a loss of focus event were typically characterized by either the first of two quick consecutive copy events followed by a single loss of focus event, or single copy events not connected temporally with a loss of focus event.  

\begin{table}
\caption{Number of copy events detected in the introductory and upper-division populations.  For reference, the total number of valid responses in the introductory and upper-division data sets was $N=1287$ and $N=308$ respectively. }\label{tab:copy}
 \begin{ruledtabular}
    \begin{tabular} {p{4cm} c c}
		& Introductory & Upper-division \\
		\hline
		Number of students with  	&	147	& 22 	\\
		\hspace{2mm} copy events & & \\
		Median number of copy	&	4	& 2 	\\
		\hspace{2mm} events per student & & \\
		Max number of copy events 	&	54	&	11	\\
		\hspace{2mm} per student & & \\
		Total number of copy events	&	861	&	67	\\
	\end{tabular}
 \end{ruledtabular}
\end{table}

Given this pattern, we also examined whether the students with copy events had any difference in performance from other students.  For the introductory RBAs, students with copy events scored higher than students without copy events (average z-score difference of $0.45$).  This trend was exactly flipped for the upper-division RBAs where students with copy events scored lower (average z-score difference of $-0.46$).  This difference was statistically significant in both cases (Mann-Whitney U $p<0.001$) and of moderate effect size (Cohen's $|d|=0.46$ in both cases).  

In the second semester of data collection, in which all data from the introductory RBAs were collected, additional Javascript code was included; this code collected information not only on when students copied text, but also from which question prompt they copied that text.  We used this information to determine if a student who copied the text of an item was more likely than the rest of the students to get that specific question correct.  To determine this, we looked at each question individually and counted how often a student who copied text from that question got it correct vs. got it incorrect.  Similarly, we counted how often students who had not copied text from that question got it correct vs. incorrect.  The result was a $2x2$ contingency table with columns denoting whether or not the student copied text from that question and rows denoting whether the student got the question correct or not.  We then summed the tables across all questions and the resulting contingency table is given in Table~\ref{tab:contT}.  

\begin{table}[b]
\caption{Contingency table breaking down how often students' responded to a question correctly relative to whether they had copied text from that question.  This table includes data from all questions; thus, each count in the table represents a response from one student to one question.   }\label{tab:contT}
 \begin{ruledtabular}
    \begin{tabular} {l c c}
		& Copied text & Did not copy text \\
		\hline
		Correct response 	&	559	& 28,291 	\\
		Incorrect response	&	163	& 20,596 	\\
	\end{tabular}
 \end{ruledtabular}
\end{table}

Table~\ref{tab:contT} shows that, on average, when introductory students copied text from an item they responded correctly to that item 77\% of the time. Alternatively, introductory students who did not copy text from a particular item responded correctly to that item only 58\% of the time, on average.  This difference in frequency is statistically significant (Chi-squared $p<<0.001$).  This shows that students who copied text from a question were more likely to get that question correct than students who did not.  We can also look at whether a student who copies text from one or more questions scores higher, on average, on those questions than on the subset of questions from which they did not copy text.  To determine this, we focused just on the $N=147$ introductory students who had one or more copy events.  We then calculated z-scores for their performance on the subset of questions where they copied text and z-scores for their performance on the subset of questions where they did not copy text.  We then average the two resulting scores across all students to determine whether students perform better on average on questions where they copied text relative to questions where they did not.  We found that the z-score on the subset of copied questions was higher on average by just under half a standard deviation (i.e., a z-score difference of 0.44).  This difference is statistically significant (Mann-Whitney U $p<<0.001$) and of moderate size (Cohen's $d=0.4$).  

Together, these results together suggest that, in the introductory courses, roughly 10\% of students do try to look up the answers to the RBA and that doing so appears to improve their performance.  In a high-stakes testing environment, this trend would be extremely problematic as it would imply that an individual students' score could not be reliably interpreted.  However, RBAs within PER are intended to be low-stakes measures of group (rather than individual performance); it is widely considered inappropriate for a range of theoretical and practical reasons to use RBAs as a measure of individual student performance \cite{engelhardt2009ctt}.  So the question then becomes, what impact does this copying behavior have on the average score for the class as a whole and, thus, our ability to interpret and compare across online and paper-based administrations of the RBA.  To determine this, we compare the average score for the full introductory data set relative to the overall score for just the subset with no copy events.  We examine this both for the total score and the scores for individual items.  

Removing all students who had copy events from the introductory data set resulted in a drop in overall average score of roughly 1\%.  This difference represents a very small effect (Cohen's $d=0.05$) and was not statistically significant (Mann-Whitney U $p=0.2$).  This suggests that the impact which looking up answers to the RBA online had on the overall course average was statistically and practically negligible.  By individual item, the difference in average item score generated by removing students who copied that item had a range of [-0.27\%, +1.4\%] with a mean of 0.29\%, suggesting that the impact of students looking up answers on individual item scores is, in practice, negligibly small.  

\subsection{\label{sec:time}Time to Completion}

We also examine the total amount of time to completion for each student to determine whether student's scores are related to how long it took them to complete the assessment.  Total time data are calculated by comparing the recorded time when the student first opened the survey link to when they made their final submission of the survey.  This does not remove periods when browser focus was lost, and can even include a period when the survey window was closed and later reopened.  As such, these duration do not necessarily reflect the amount of time a student actually worked on the assessment, merely the amount of time that passed between them opening and submitting the assessment.  For the vast majority of students (65\%, $N=843$ of 1287 in introductory; 78\%, $N=239$ of 308 in upper-division), the total time between start and submit fell within a time frame of 15-60 min, consistent with what would be required of a student taking the RBA in class.  Total time spent on the RBA showed a significant (though small) correlation with z-score on the assessment only for the introductory students (Spearman $r=0.3$, $p<<0.001$).  

We can also use the focus data to modify the raw time data by subtracting out the total time for each student during which their survey window was hidden, suggesting they may not have been working on the assessment.  Doing so does not significantly shift either the number of students whose total time (now excluding time away from the browser) falls between 15-60 min or the correlation of time with score on the assessment for either introductory or upper-division students.

\section{\label{sec:discussion}Discussion \& Limitations}
We collected online responses to four research-based assessments spanning both introductory and upper-division content.  This work is part of ongoing research to determine whether students' performance on RBAs shifts when these assessments are given online.  For three of the courses in the data set, we also have historical scores from students in these same classes with the same instructor where the RBA was given on paper and during class.  Comparisons of the online and in-class scores showed the online scores being roughly 5\% lower.  This difference was statistically significant only in the case of the introductory population (two-tailed t-test, $p=0.001$) though the effect was small (Cohen's $d=0.13$).  The decrease in average score appeared to be largely driven by the presence of a larger tail of the distribution (in terms of grades) in the online administrations.  This, combined with the higher participation rates in the online administration suggests that administering RBAs online to an upper-division population encourages more of the lower performing students to participate.  

In addition to students' responses to the RBAs, we also collected data using embedded JavaScript code on students' online behaviors such as copying text, printing browser pages, and loosing browser focus by clicking into other browser tabs.  We found that only a small number of students (less than 0.5\%) printed or copied item prompts in a manner that suggested they were attempting to save some or all of the item prompts.  Such behavior primarily represents a potential concern with respect to test security if students chose to post the assessment prompts online.  However, we have anecdotal evidence that at least some of these students were saving the prompts solely for their own future studying and with no intention of sharing them.  How much of a concern maintaining test security is may also vary between introductory and upper-division RBAs.  Our own attempts to look up solutions to the RBAs used in this study showed that item prompts and solutions to the FMCE and BEMA are already available online on paid solution sites.  Alternatively, we found no evidence of item prompts or solutions for the CUE and QMCA.  Thus, test security has already been at least partially breached for the introductory RBAs, but appears to be largely intact for the upper-division RBAs.  This may be a reflection of the fact that the FMCE and BEMA are both older and more widely used assessments than the CUE and QMCA.  


We also collected data on how often and for how long students clicked out of their RBA browser tab as a proxy for distraction.  Such behavior was common, with roughly half the students engaging in online behaviors resulting in loss of browser focus and indicating that the students may have disengaged from the RBA for a period of time.  However, roughly two-thirds of the periods where students lost browser focus lasted less than 1 minute, and less than 10\% of the periods lasted for longer than 5 minutes.  The total amount of time spent away from the assessment tab had a small but statistically significant positive correlation with overall score on the RBA only for the introductory population.  Thus, we argue that while the potential for distraction and disengagement certainly increases with online RBAs, our data suggest the majority of students do not become disengaged for long periods and that this disengagement does not appear to negatively impact their performance.  The slight positive correlation that appears for the introductory students is unexpected when considering time away from the assessments tab as a proxy for disengagement.  This trend may be driven by students who navigated away from their RBA browser tab when accessing internet resources to assist them in completing the assessment.  

Evidence of copying text was observed in roughly ten percent of the students in our sample.  Roughly three-quarters of these copy events were immediately followed by a browser focus event in which the RBA tab became hidden.  Such a pattern is consistent with what we would observe if students were attempting to Google the item prompts in an attempt to determine the correct answers.  While it is not possible for us to determine for certain if that is what the students were doing, the pattern is suggestive.  Moreover, students with copy events had statistically different score distributions than the rest of the populations.  However, the trend differs between the introductory and upper-division students.  Upper-division students with copy events scored lower than other students, while introductory students scored higher.  Using information on which specific questions students copied text from, we found that students who copied text for a particular question more often got that question correct. Moreover, we found that students scored, on average, higher on the subset of items from which they copied text than from the items they did not.  

Taken together, the findings summarized above are consistent with the follow interpretation.  A small subset of students do attempt to Google item prompts when taking online RBAs, and, as evidenced by the lower performance of these students in the upper-division population, these students may deferentially include lower performing students.  In cases where the solutions to the specific RBAs are not easily accessible online (e.g., the CUE and QMCA), copying and Googling text does not improve students scores.  Alternatively, in cases where the solutions to the specific RBA are available online, copying and Googling text does result in an improvement in students performance.  However, because the improvement to students scores is, on average small (roughly a third of a standard deviation of improvement in average score) and impacts only a small fraction of students, the impact of this behavior on class average scores overall or by question was negligibly small.  

Overall, our findings suggest that while students in our sample engaged in a variety of online behaviors, none of these behaviors resulted in a change in the population's average performance that would threaten our ability to interpret this performance or compare it to paper-based implementations of the RBA.  However, this held largely because only a small component of the student population actually engaged in some of these behaviors.  Should the number of students engaging in, specifically, copying behaviors increase in the future, the impact of these behaviors on the course average may increase.  

The only effect observed in the current study that presents a concern for comparisons of online and paper-based RBAs was a consistent roughly 5\% drop in overall average score relative to historical paper-based implementations in these courses.  This drop may be a result of the larger participation rates observed in the online administrations and, thus, the inclusion of a larger component of the lower performing tail of the student population.  Rather than being a problem, we argue this actually represents an advantage for the online RBAs in that they appear to provide a broader sample of the student population.  

The work presented here has some important limitations.  The code that captured students' online behaviors can only detect actions at the browser level, meaning that actions at the computer level (like switching into a new program) cannot be detected.  For this reason, our data should be interpreted as a lower bound on the appearance of these behaviors.  Replication of this work with additional RBAs, with a broader student population, and in future semesters will be important to ensuring that these results hold across different tests, a broad student population, and time.  However, these results do suggest that online assessment is a promising alternative that brings with it many potential logistical advantages.

\begin{acknowledgments}
This work was funded by the CU Physics Department.  Special thank you to the faculty and students who participated in the study and the members of PER@C for all their feedback.  
\end{acknowledgments}

\bibliography{master-refs-01-19}

\end{document}